\documentclass[traditabstract]{aa} 
\usepackage{graphicx}
\usepackage{txfonts}
\begin{document}
\title{Optical depths for gamma-rays in the radiation field of a star heated 
       by external X-ray source in LMXBs}

   \subtitle{Application to Her X-1 and Sco X-1}

   \author{W. Bednarek \& J. Pabich
          }

 \institute{Department of Astrophysics, University of \L \'od\'z,
              90-236 \L \'od\'z, ul. Pomorska 149/153, Poland\\
             \email{bednar@astro.phys.uni.lodz.pl}}

   \date{Received ; accepted }


\abstract{The surface of a low mass star inside a compact low mass X-ray binary system (LMXB) can be heated by the external X-ray source which may appear due to the accretion 
process onto a companion compact object (a neutron star or a black hole). As a result, the surface temperature of the star can become significantly higher than it is in the normal state resulting from thermonuclear burning.
We wonder whether high energy electrons and gamma-rays, injected within the binary system, can efficiently interact with this enhanced radiation field. To decide this, we calculate the optical depths for the gamma-ray photons in the radiation field of such irradiated star as a function of the phase of the binary system. Based on these calculations, we conclude that compact low mass X-ray binary systems may also become sources of high energy gamma-rays since conditions
for interaction of electrons and $\gamma$-rays are quite similar to these ones observed within the high mass TeV $\gamma$-ray binaries such as LS 5039 and LSI 303 +61.
However, due to differences in the soft radiation field, the expected $\gamma$-ray light curves can significantly differ between low mass and high mass X-ray binaries.  
As an example, we apply such calculations to two well known LMXBs: Her X-1 and Sco X-1.
It is concluded that electrons accelerated to high energies inside these binaries should find enough soft photon target from the companion star for efficient $\gamma$-ray production.
}

\titlerunning{Optical depths for gamma-rays in LMXBs}

\maketitle

\section{Introduction}

Recently TeV $\gamma$-rays have been observed from a few massive X-ray binaries (e.g.~Aharonian et al.~2005a, Albert et al.~2006). 
Up to now such $\gamma$-ray emission has not been reported from the low mass X-ray binaries (see e.g. observations of SS 433, Aharonian et al.~2005b), which are very luminous X-ray sources ($L_{\rm X}\sim 10^{36-39}$ erg s$^{-1}$, see the catalog of LMXB~Liu et al. 2007).
However, the class of LMXBs as a whole has not been monitored intensively by the modern Cherenkov telescopes up to now. In this paper we try to argue that very compact LMXBs are also promissing targets for TeV $\gamma$-ray astronomy.

LMXBs likely contain a neutron star which accretes matter from a companion star. Sometimes X-ray emission shows also pulsations with the rotational period of the neutron star~(Nagase~1989, Karino~2007). Some LMXBs show also optical modulation with the period of the binary system  which can be interpreted as a result of external heating of the stellar surface by the X-ray source (e.g. Her X-1, Sco X-1).

The mechanism for particle acceleration and $\gamma$-ray production in the massive TeV $\gamma$-ray binaries is not at present clear. Two general scenarios are considered.
In the first one, energetic pulsar creates relativistic wind which collides with the strong wind of the massive star creating a shock wave (e.g. Maraschi \& Treves~1981, Vestrand \& Eichler~1982, Harding \& Gaisser~1990, Tavani \& Arons~1997, Kirk et al.~1999). Electrons can be accelerated up to 
the TeV energies in the shock acceleration scenario or in the not well known mechanism occurring in the pulsar wind. In the second scenario, a compact object accretes  matter from the massive star. Electrons can be accelerated in the shock waves created
in the stream of matter expelled from the inner part of the accretion disk (a jet) 
(e.g. Levinson \& Blandford~1996, Georganopoulos et al.~2002, Romero et al.~2002) or at the border in which accreting matter is stopped by the rotating neutron star magnetosphere (Bednarek~2009a,b). The second scenario can also occur in the case of low mass X-ray binaries. Therefore, it is possible that acceleration of particles to TeV energies should also occur within the low mass X-ray binaries. 
Note that TeV $\gamma$-ray emission is observed from variety of cosmic sources (jets of active galaxies, vicinity of pulsars, supernova remnants, clusters of stars) which significantly differ in structure and physical parameters. So then, even if parameters of 
objects responsible for acceleration of particles within HMXBs and LMXBs may differ significantly,
this does not exclude acceleration of particles to TeV energies in LMXBs. However the efficiency
of particle acceleration may differ. The main purpose of searching for $\gamma$-ray emission from LMXBs with the present and future Cherenkov telescopes (e.g. CTA) will be to put light on this basic question. The aim of this paper is to show what are the most favourable conditions for
detection of TeV $\gamma$-ray signal from the class of very compact LMXBs.

Efficient $\gamma$-ray production within massive binaries likely occurs due to the comptonization by relativistic electrons of the soft radiation produced by the massive star. Such strong radiation is not present inside the low mass binaries where a normal star has characteristic surface temperature below $\sim 10^4$ K. Therefore, in general, efficient TeV $\gamma$-ray production is not expected there. 
However, in very compact LMXBs, a low mass star can be extensively heated by the X-ray source produced due to accretion of matter onto companion compact object. As a result, some parts of the stellar surface can reach temperature comparable to those observed in the case of high mass stars (i.e. a few $10^4$ K). This soft radiation provides enough
target for relativistic electrons, accelerated inside such compact low mass binaries, for production of TeV $\gamma$-rays. In this paper we calculate the optical depths for $\gamma$-ray photons in the radiation field of irradiated low mass star in order to check whether $\gamma$-ray production in such systems can occur efficient and whether produced $\gamma$-rays can escape from the vicinity of the star. Note, that the cross sections for $\gamma$-ray production in the inverse Compton scattering (ICS) and $\gamma$-ray absorption in collision with soft photons are comparable. Therefore,
directions within the binary system in which significant absorption of $\gamma$-rays can occur should also correspond to directions of efficient $\gamma$-ray production in ICS process.

\begin{figure}
\vskip 9.5truecm
\includegraphics{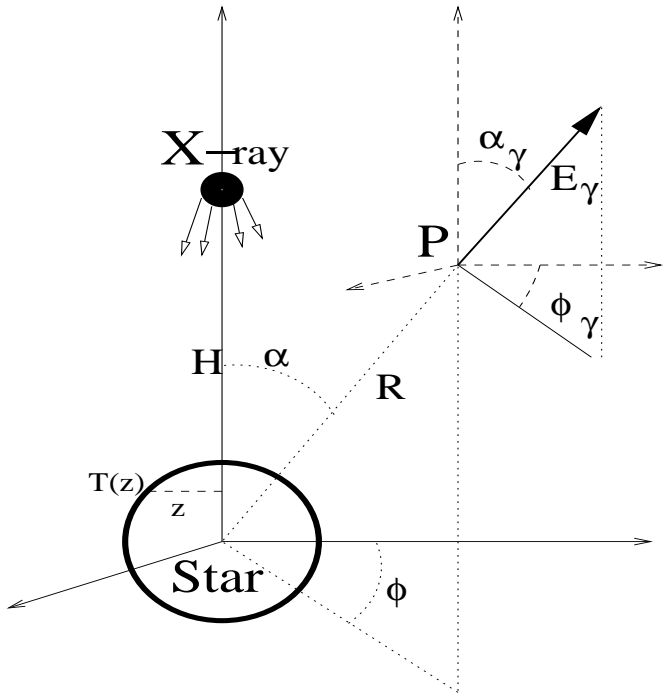}
\caption{Schematic picture of the geometrical situation considered in the paper.
$\gamma$-ray photon with energy $E_\gamma$ is injected from a place at the distance $R$ from the center of the star and a direction defined by the angle $\alpha$ and $\phi$. $\gamma$-ray propagates at the angles $\alpha_\gamma$ and $\phi_\gamma$.
The star is irradiated by a point X-ray source (with the luminosity $L_{\rm X}$) which is at the distance $H$. As a result of irradiation, a temperature gradient appears on the stellar surface (T(z)) with the maximum value at the closest point to the X-ray source. }
\label{fig1}
\end{figure}
\section{Soft radiation from irradiated star}

The X-ray emission produced in the vicinity of a compact source in some close low mass X-ray binaries can reach the values of $\sim 10^{36-39}$ erg s$^{-1}$. A part of this X-ray emission can illuminate the surface of a companion star. As a result, the surface temperature rises significantly. We calculate the temperature profile onto the surface of such irradiated star by
a point X-ray source located at the distance, $H$, from the center of the star. 
We assume that all X-ray emission absorbed by the stellar surface is irradiated as a black body emission with characteristic temperature $T(z)$, where $z$ is the distance measured
from the direction defined by the center of the star and the X-ray source and a specific point onto the stellar surface (see Fig.~1). Simple considerations
turn out to the temperature profile on the stellar surface of the type,
\begin{eqnarray}
T(z)  = \left(T_\star^4 + {{L_{\rm X}\cos\beta/4\pi \sigma_{\rm SB}R_\odot^2}\over{z^2 + 
\left(H - \sqrt{R_\star^2 - z^2}\right)^2}}\right)^{1/4}\nonumber\\
   \approx   
7.3\times 10^4 \left({{L_{38}\cos\beta}\over{z^2 + \left(H -\sqrt{R_\star^2 - z^2}\right)^2}}\right)^{1/4} {\rm K},
\label{eq1}
\end{eqnarray} 
\noindent
where distances are defined in Fig.~1, $L_{\rm X} = 10^{38}L_{38}$ erg s$^{-1}$ is the luminosity of the X-ray source, $\sigma_{\rm SB}$ is the Stefan-Boltzmann constant, 
$\beta = \pi - \mu - \nu$, $\cos\mu = z/R_\star$, $\tan\nu = \left(H - \sqrt{R_\star^2 - z^2}\right)/z$, and $R_\star$ is the stellar radius.
The distances $H$, $R$, $R_\star$, and $z$ are expressed in units of the Solar radius $R_\odot = 7\times 10^{10}$ cm.
The value of $z$ can change in the range from 0 to $z_{\rm max} = R_\star\sqrt{H^2 - R_\star^2}/H$.

\begin{figure*}
\vskip 14.3truecm
\includegraphics{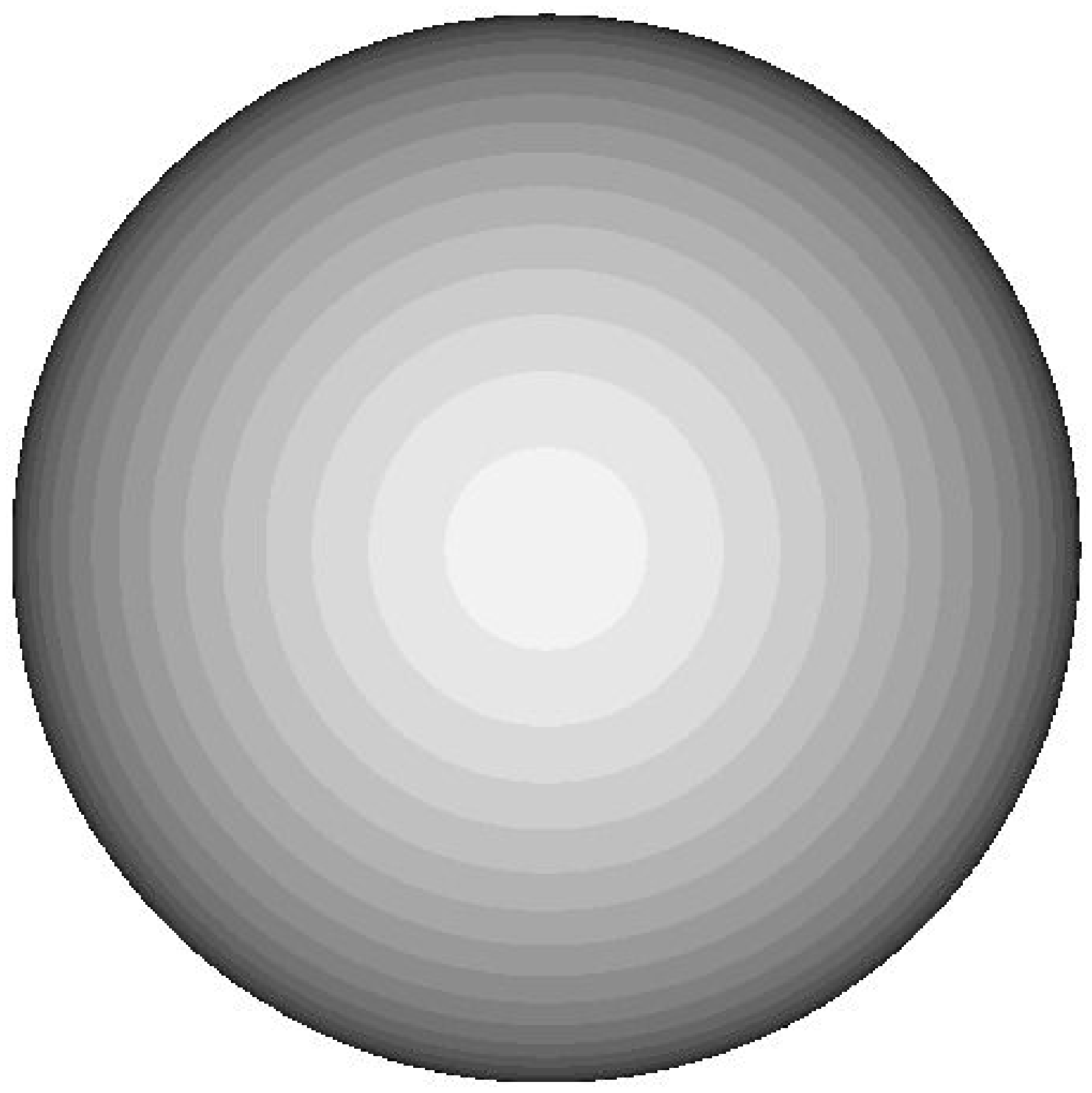}
\includegraphics{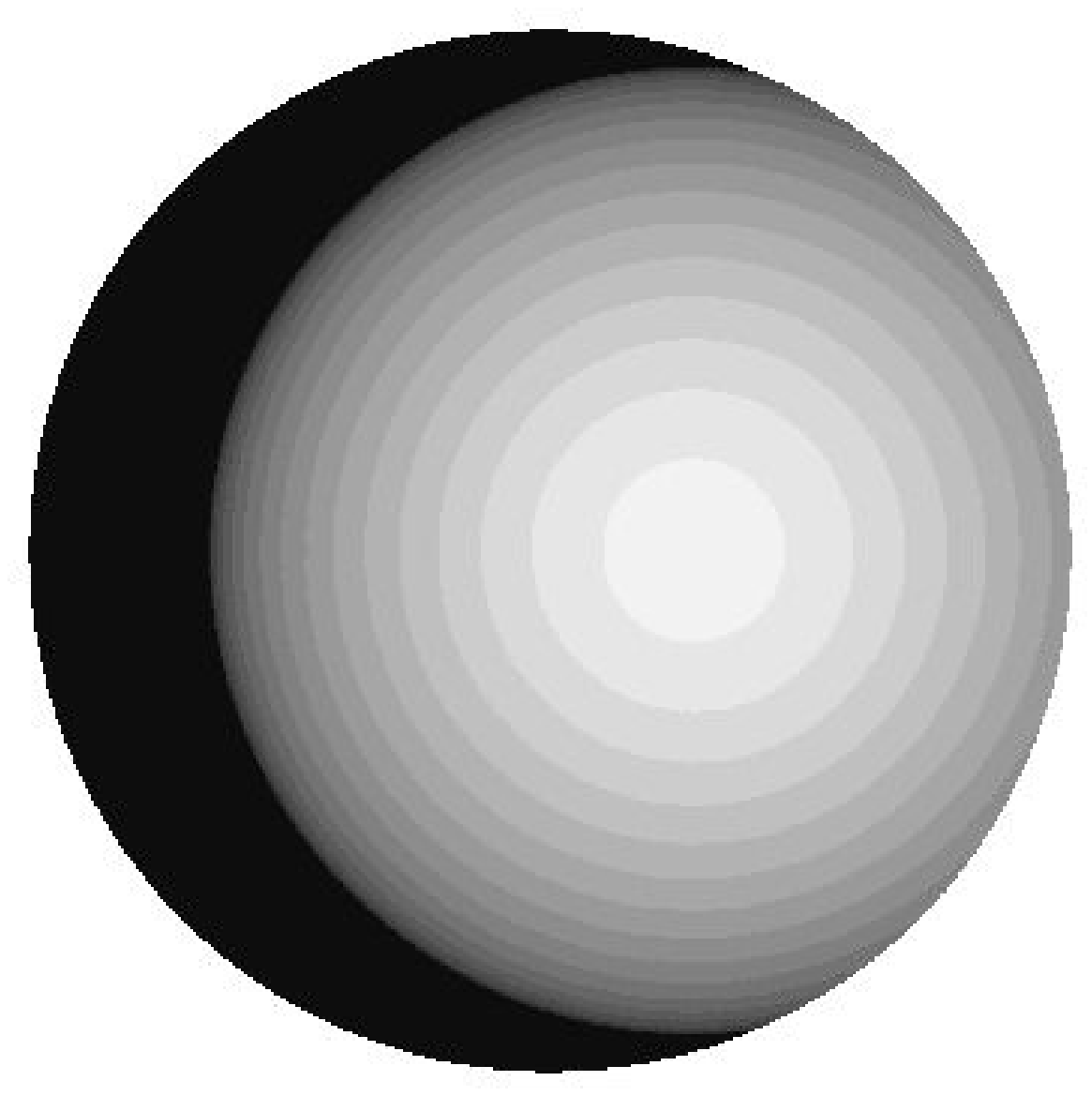}
\includegraphics{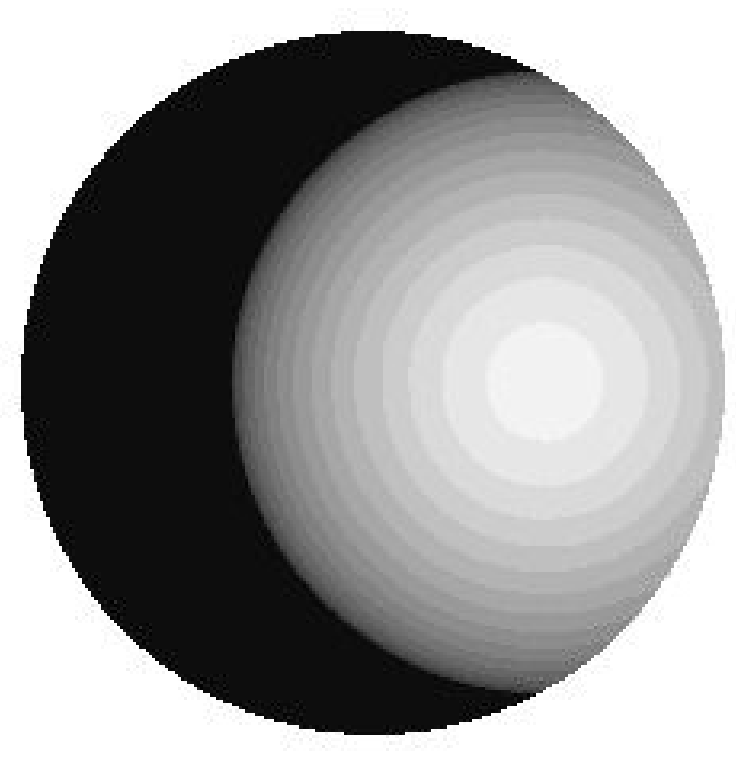}
\includegraphics{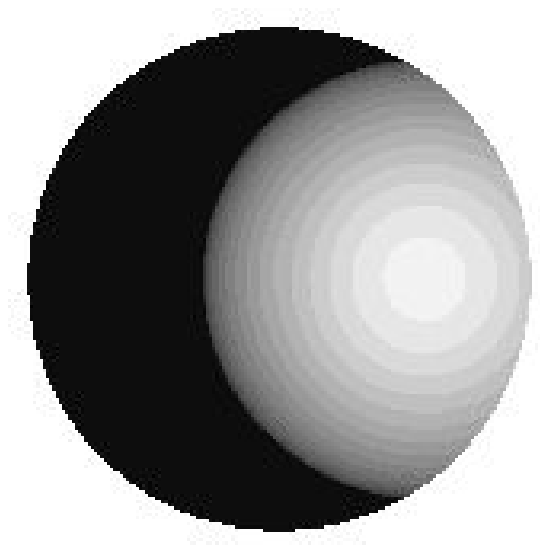}
\includegraphics{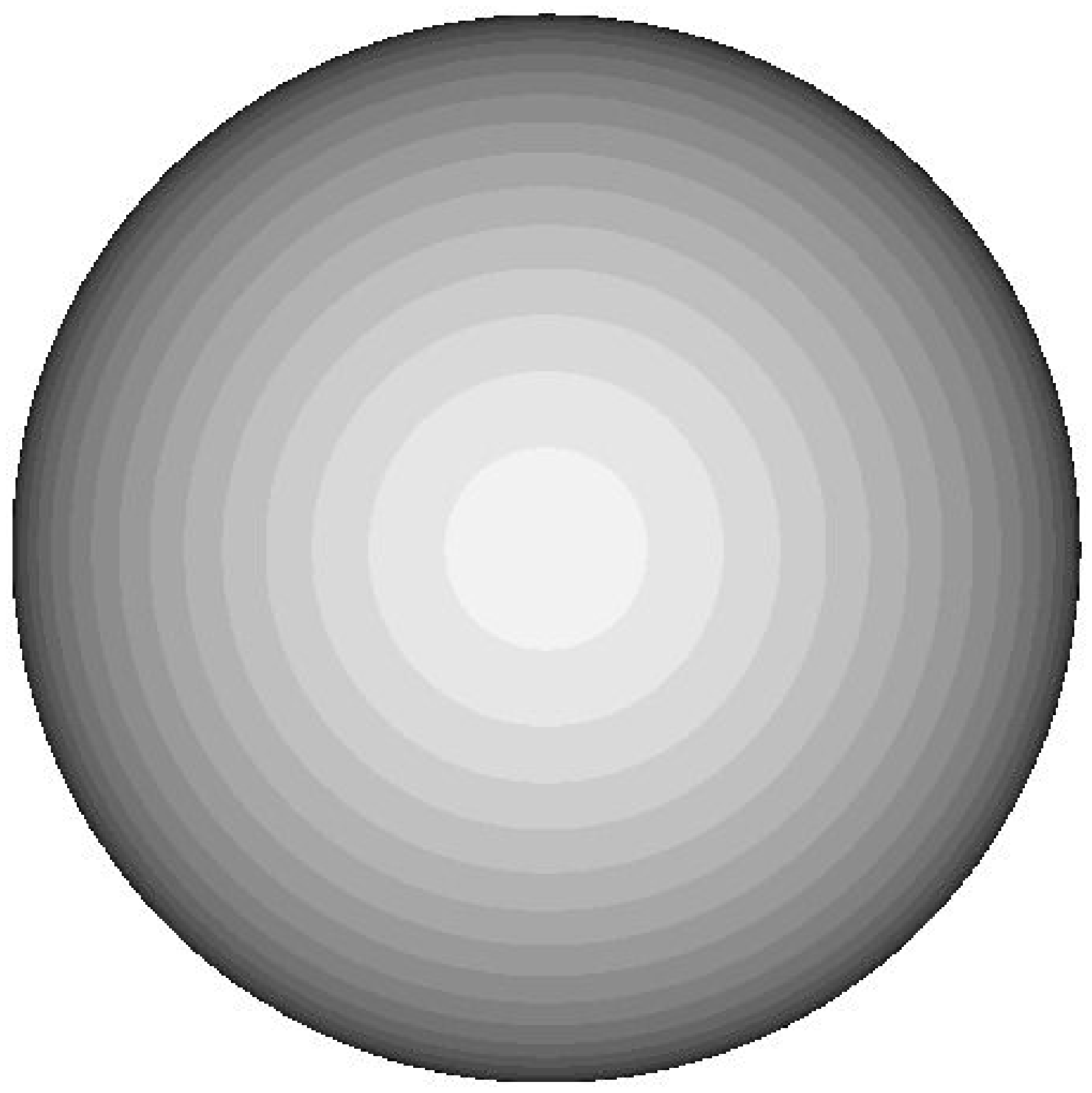}
\includegraphics{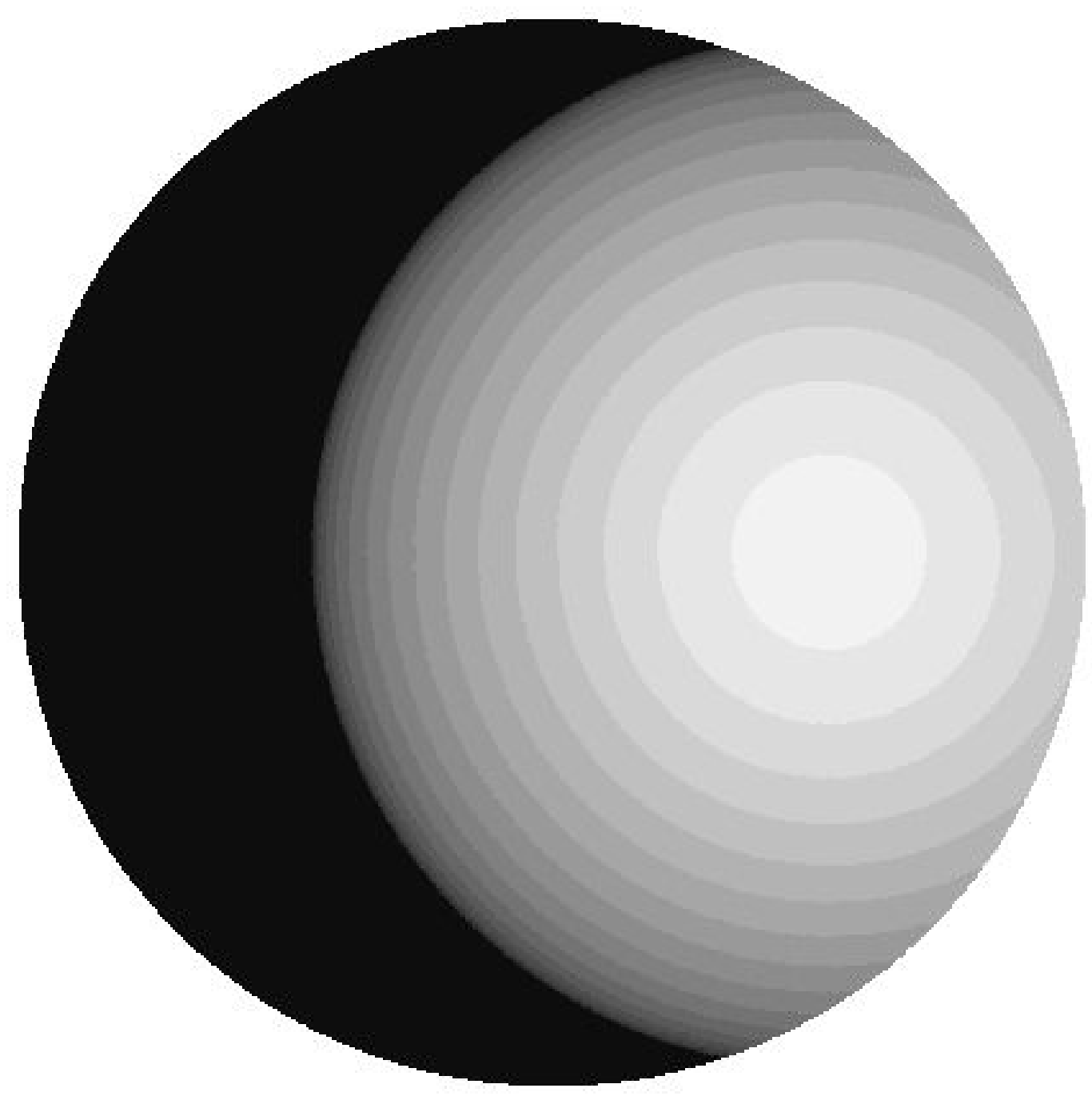}
\includegraphics{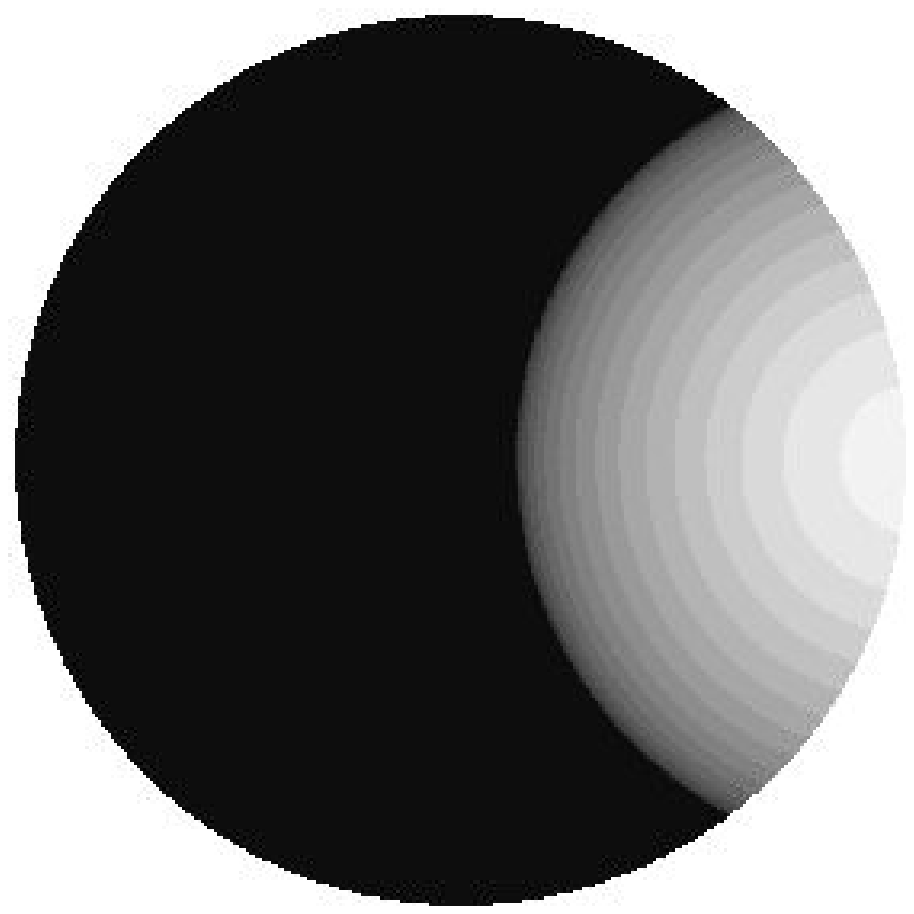}
\includegraphics{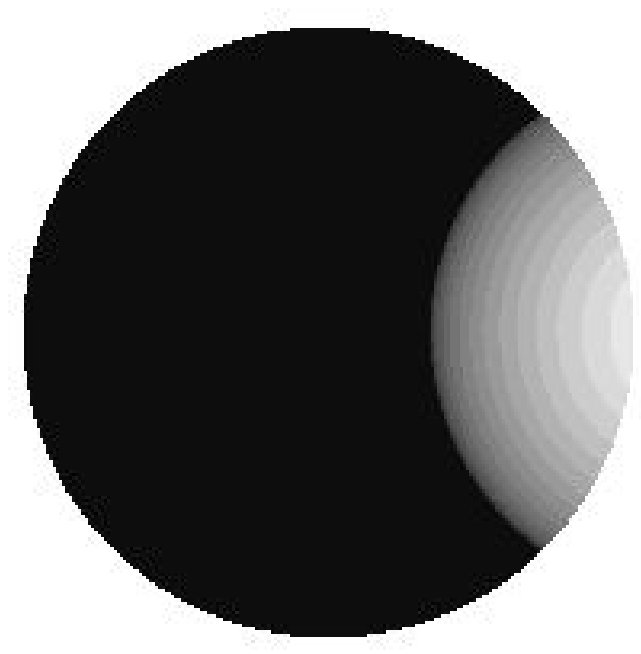}
\includegraphics{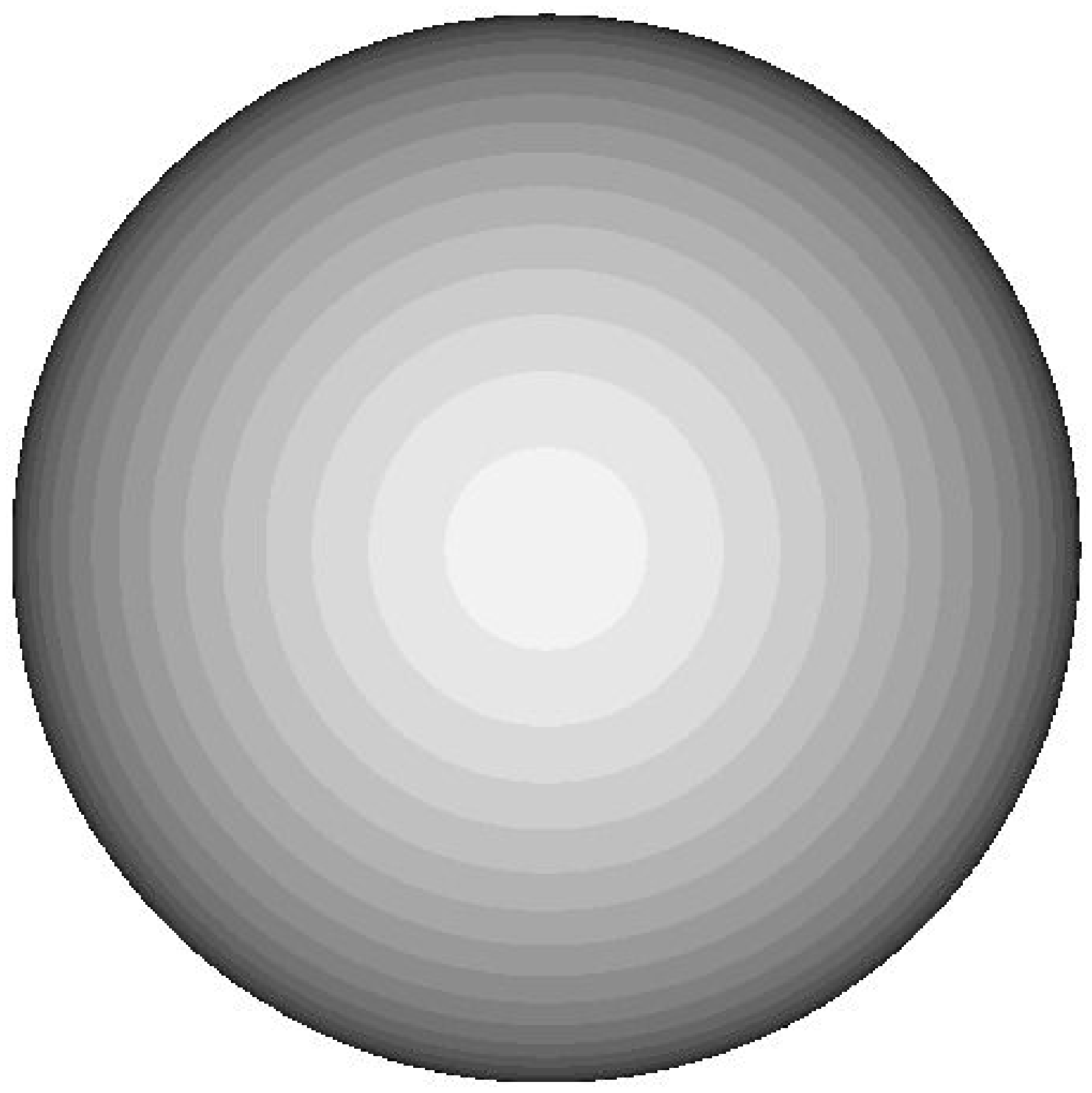}
\includegraphics{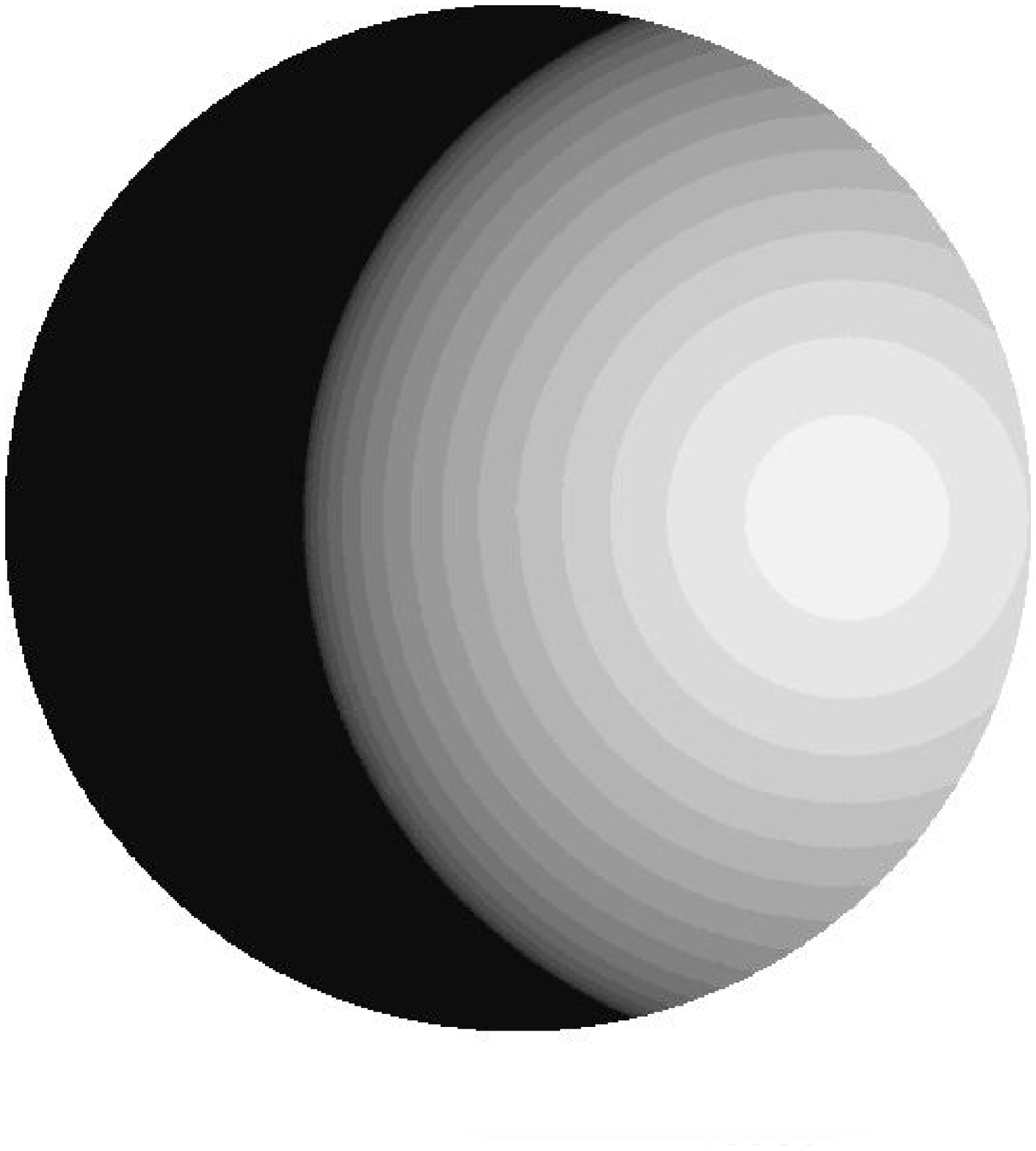}
\includegraphics{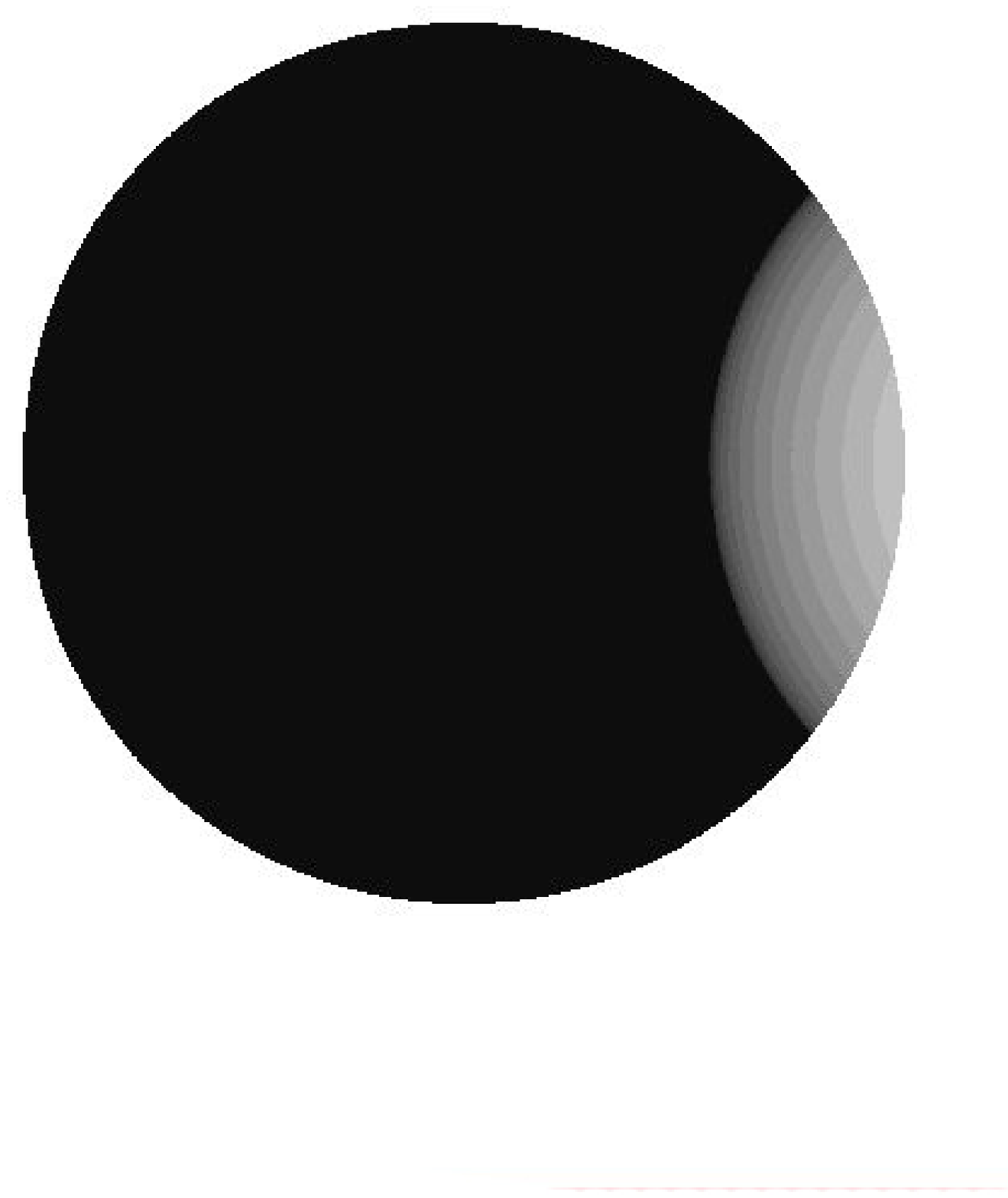}
\includegraphics{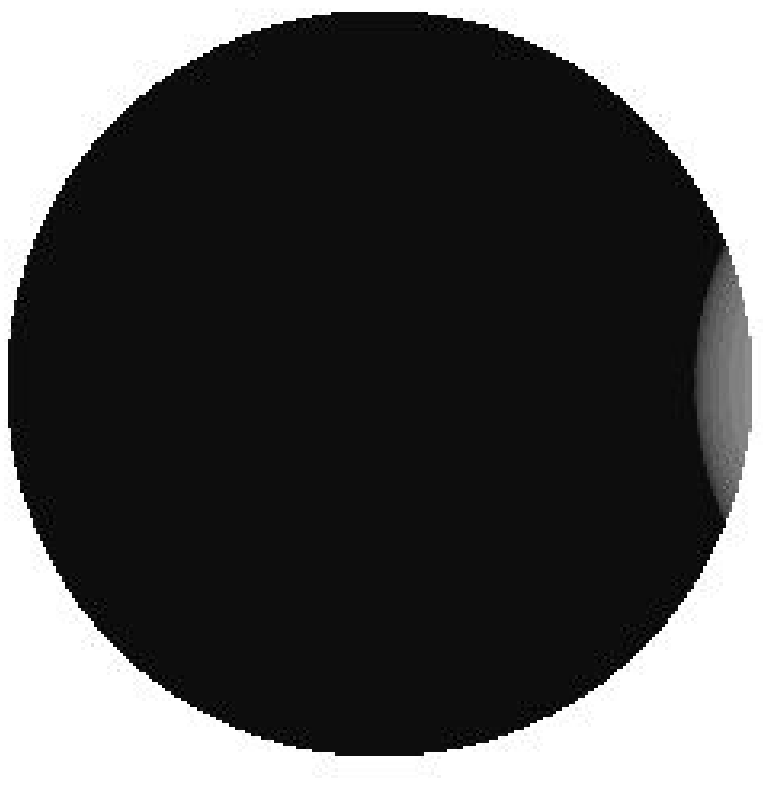}
\caption{The radiation field seen by $\gamma$-ray photon at its different locations along the direction of propagation $L$. This radiation is due to the irradiation by external X-ray source with luminosity $10^{38}$ erg s$^{-1}$ located at the distance 
$H = 2R_\star$ from the center of the star.
$\gamma$-rays are also injected at the distance $H = 2R_\star$ but at different angles $\alpha_\gamma = 45^{\rm o}$ (upper panel), $90^{\rm o}$ (middle panel), and $120^{\rm o}$ (bottom panel) in respect to the line defined by the injection point and the center of the star.
The propagation distance is equal to $L = 0$ (left), $R_\star$ (left-middle), $3R_\star$ (right-middle), and  $5R_\star$ (right). The star has the radius $R_\star = R_\odot$ equal to the radius of the Sun and the surface temperature $T_\star = 6000$ K. The grey scale denotes the range of temperature starting from
$7.5\times 10^4$ K (white) to $6\times 10^3$ K (black).}
\label{fig2}
\end{figure*}

The distribution of temperature on the surface of such irradiated star seen by the external observer located at different places outside the star is shown in Fig.~2. 
Two effects determine this soft radiation field at a specific place: (a) the distance,
$R$, from the stellar surface determines the solid angle intercepted by the star; (b) the angle, $\alpha$, at which the hot region is seen (measured in respect to the direction defined by the X-ray source and the star).
Note, that for reasonable parameters of the low mass X-ray binary system (e.g. $L_{\rm X} = 10^{38}$ erg s$^{-1}$, and $H = 2 R_\star$), the surface temperature can increase due to irradiation of the star up to a few $10^4$ K. This is comparable to the
surface temperature of the massive stars within the high mass X-ray binaries. Therefore, we expect that the optical depths for TeV $\gamma$-rays in the soft radiation of such irradiated low mass stars can also reach substantial values, exceeding unity in some cases. 

Since the $\gamma$-ray photon injected at a specific place sees the hot part of the star at different angles, the soft radiation field seen by the $\gamma$-ray photon changes significantly during its propagation in the vicinity of the star (due to the change of the viewing angle and the distance from the stellar surface). In Fig.~2 we show the example radiation field from the stellar surface as seen by the $\gamma$-ray photon at different propagation distances $L$ assuming that the primary $\gamma$-ray is injected at the X-ray source. Note that, for the propagation angles $\alpha_\gamma > 90^{\rm o}$ (see Fig.~1), the solid angle subtracted by the star at first increases since the $\gamma$-ray approaches the stellar surface. For small angles $\alpha_\gamma$, the solid angle subtracted by the star continuously drops with the $\gamma$-ray propagation path and the hot region becomes less visible.

\section{The optical depths for gamma-rays}

The optical depths for $\gamma$-ray photons injected at an arbitrary distance from the surface of the star in the case of its fixed surface temperature were calculated for the first time in the general case by Bednarek~(1997,2000). In respect to previous calculations (e.g. Moskalenko et al.~1993), Bednarek took also into account dimensions of the star, which allows their application to very compact binaries such as e.g. Cyg X-3. Such calculations are necessary in the case of the IC $e^\pm$ pair cascades developing within the massive binaries since in principle secondary $e^\pm$ pairs and $\gamma$-rays can appear everywhere within the binary. They can also fall onto the surface of the companion star. The optical depths for $\gamma$-rays calculated in the radiation field of the star with specific parameters can be easily re-scaled for the cases of massive stars with other surface temperatures and radii (Bednarek~2009a). For example, $\gamma$-ray photons with energies, $E_\gamma^{\rm o}$, propagating at specific distance $D$ from the star, and in direction (defined by the angle $\alpha$), approaching close to the star with specific parameters ($T_{\rm o}$ and $R_{\rm o}$) are related to the optical depths around arbitrary stars with $T_\star$ and $R_\star$ in the following way,
\begin{eqnarray}
\tau({{E_\gamma^{\rm o}}\over{S_{\rm T}}},T_\star,R_\star,D,\alpha)
= S_{\rm T}^3S_{\rm R}\tau(E_\gamma^{\rm o}, T_{\rm o}, R_{\rm o},D,\alpha)
\label{eq1}
\end{eqnarray}
\noindent
where $S_{\rm T} = T_\star/T_{\rm o}$, $S_{\rm R} = R_\star/R_{\rm o}$, and the distance $D$ is measured in stellar radii. These $\gamma$-ray  optical depths have been frequently discussed in the context of the massive stars inside the high mass X-ray binaries recently detected in the TeV $\gamma$-rays (see e.g. B\"ottcher \& Dermer~2005, Bednarek~2006a, Dubus~2006).

\begin{figure*}
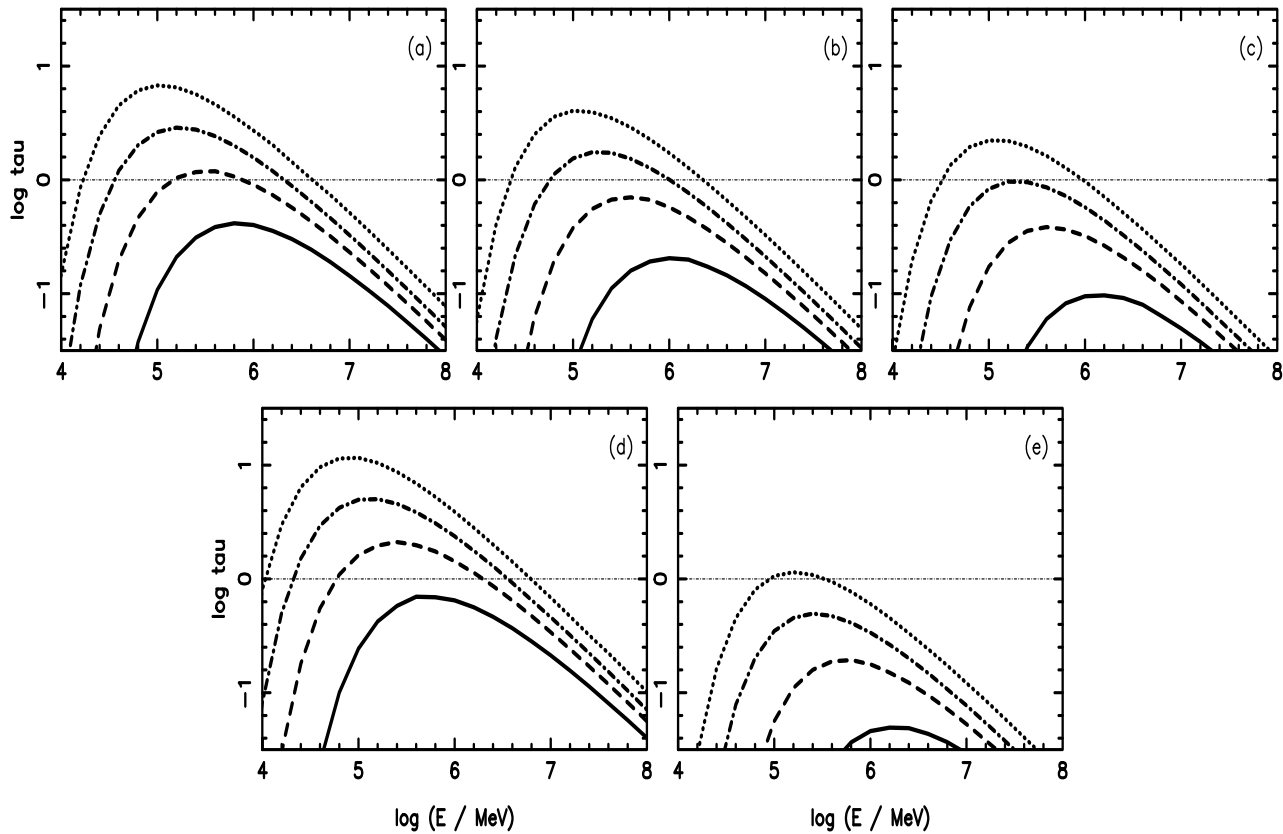

\vskip 12.2truecm
\includegraphics{bpfig3a.eps}
\includegraphics{bpfig3b.eps}
\includegraphics{bpfig3c.eps}
\includegraphics{bpfig3d.eps}
\includegraphics{bpfig3e.eps}
\caption{The optical depths for $\gamma$-rays in the radiation field created by a low mass star irradiated by external X-ray source. It is assumed that the X-ray source is at the distance $H = 2R_\star$ (upper panel) from the center of the star. The $\gamma$-rays are injected at different distances from the star (see Fig.~1): $R = 1.5R_\star$ (left figure), $2R_\star$ (middle), and $3R_\star$ (right), and the angles $\alpha = 0^{\rm o}$ and $\phi = 0^{\rm o}$. $\gamma$-ray photon moves at  different angle $\alpha_\gamma = 30^{\rm o}$ (solid), $60^{\rm o}$ (dashed), $90^{\rm o}$ (dot-dashed), and $120^{\rm o}$ (dotted). The optical depths for other distances of the X-ray source and injection places of $\gamma$-rays, $H = 1.5R_\star$ and $R = 1.5R_\star$ and  $H = 3R_\star$ and $R = 3R_\star$, are shown in the bottom panel on the left and right figures, respectively. The parameters of the star in the binary system are: $T_\star = 6000$ K, $R_\star = R_\odot = 7\times 10^{10}$ cm.}
\label{fig3}
\end{figure*}

Here we are interested in a more complicated scenario which can be appropriate for the  low mass X-ray binary systems. In LMXBs the companion stars produce relatively weak soft radiation field due to its low surface temperature resulting from nuclear burning. However, in the presence of a nearby strong X-ray source, the surface temperature of a star can significantly increase due to the irradiation process (see section~2). Therefore, relativistic electrons injected not far from the surface of the low mass star in a compact X-ray binary can also suffer strong energy losses on the ICS process. Primary $\gamma$-rays and secondary cascade $\gamma$-rays can be efficiently absorbed in this soft radiation field.
In order to check whether such processes may become important, we calculate the optical depths for $\gamma$-rays in such more complicated scenario following the standard prescription,
\begin{equation}
\tau=\int_\ell dl \int d\epsilon d\Omega n(l, \epsilon, \Omega) \sigma_{\gamma\gamma}(\epsilon,\theta)(1-\cos\theta), \label{eq:tau}
\label{eq2}
\end{equation}
where $n(l, \epsilon, \Omega)$ is the differential density of soft photons with energy $\epsilon$ which arrive from the low mass star inside the solid angle $\Omega$ to instantaneous location of the $\gamma$-ray photon at the propagation distance $l$, $\sigma_{\gamma\gamma}$ is the $e^\pm$ pair production cross section, and $\theta$ is the angle between the momentum vectors of the gamma-ray and soft photon. $\ell$ denotes the path along propagation direction of the gamma-ray photon in the soft radiation field.

\begin{figure*}
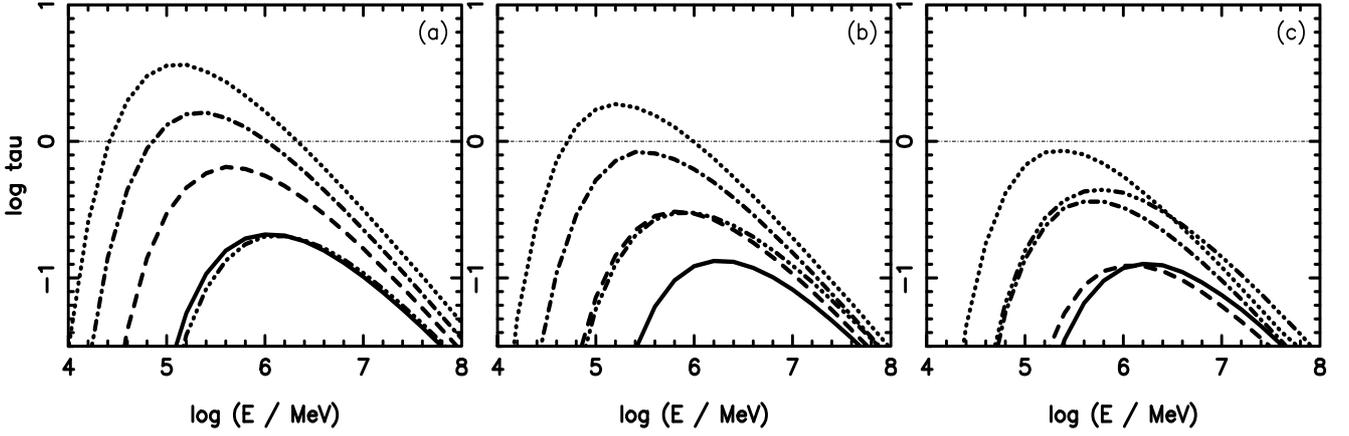

\vskip 6.2truecm
\includegraphics{bpfig4a.eps}
\includegraphics{bpfig4b.eps}
\includegraphics{bpfig4c.eps}
\caption{As in Fig.~3 but for the X-ray source at the distance $H = 2R_\star$, injection distance of $\gamma$-rays from the star $R = 1.5R_\star$
and different angles of the injection place $\alpha = 10^{\rm o}$ (left figure),
$20^{\rm o}$ (middle), and $30^{\rm o}$ (right).
$\gamma$-ray photon moves at different angle $\alpha_\gamma = 1^{\rm o}$ (triple-dot-dashed curve), $31^{\rm o}$ (solid), $61^{\rm o}$ (dashed), $91^{\rm o}$ (dot-dashed), and $121^{\rm o}$ (dotted).}
\label{fig4}
\end{figure*}

We investigate the optical depths for $\gamma$-rays as a function of their energies and other free parameters describing the geo\-me\-try of the picture, such as the injection distance $R$ and the angle $\alpha$ (see Fig.~1). The low mass stars with different radii and surface temperatures are considered. For the purpose of the example calculations, the X-ray luminosity of the compact object (a neutron star) is fixed on $L_{\rm X} = 10^{38}$ erg s$^{-1}$. The optical depths as a function of $\gamma$-ray photon energy for selected injection angles and distances from the companion star are shown in Fig.~3.
The calculations show that the optical depths are sig\-ni\-fi\-cant for some range of investigated energies of $\gamma$-ray photons. They are clearly above unity provided that $\gamma$-rays are injected within $R\sim 2R_\star$ from the center of the star at a part of the hemisphere containing the star. Note that the optical depths increase also for stars with lower radii in the case of similar compactness of the binary system (expressed as the ratio of the injection distance and the radius of the star). This is due to the fact that irradiating X-ray source is closer to the stellar surface in such cases. Note that the amount of power supplied to the stellar surface from the X-ray source increases as a square of its distance but the optical depths depend only linearly with this distance. Due to the closer X-ray source, the surface is heated to higher temperature.

In Fig.~4, we show how the optical depths depend on the angle $\alpha$ (which is the angle between the direction defined by the location of the X-ray source and the center of the star and the injection place of $\gamma$-rays and the center of the star, see Fig.~1). Such situation may correspond to the case of injection of $\gamma$-rays from different places of the jet launched from the compact object. Note, that for larger angles $\alpha$ the hottest region on the stellar surface is also seen under larger angles. Therefore, the effective radiation field from the stellar surface seen from the injection place of the $\gamma$-ray drops significantly. As a consequence the optical depths for $\gamma$-ray photons are lower (compare e.g. Figs.~4a and 4c).

In summary, we conclude that the optical depths for $\gamma$-rays can also exceed
unity in the case of very compact low mass X-ray binary systems. Therefore, we predict that these LMXBs can become sources of GeV-TeV $\gamma$-rays as in the case of high mass X-ray binaries, provided that the acceleration mecha\-nism of electrons to TeV energies still works.

\begin{figure}
\vskip 8.5truecm
\includegraphics{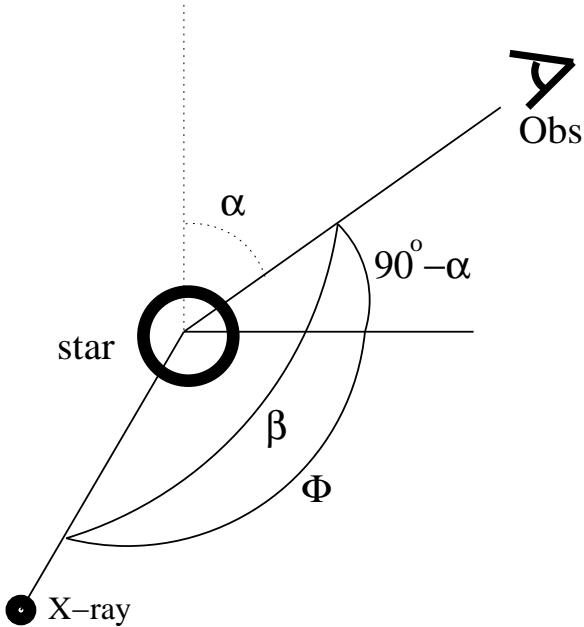}
\caption{The geometrical situation considered in the case of a compact X-ray source on an orbit around a low mass star. The X-ray source irradiates the stellar surface heating it to temperature significantly above that one due to nuclear burning. The observer sees the heated stellar surface at the angle $\beta$ measured in respect to the direction defined by the X-ray source and the star. The system is inclined at the angle $\alpha$ in respect to the observer. The X-ray source and the $\gamma$-ray source are located at the phase $\Phi$ in respect to the plane of the observer.}
\label{fig5}
\end{figure}

\section{Gamma-ray optical depth light curves}

\begin{figure}
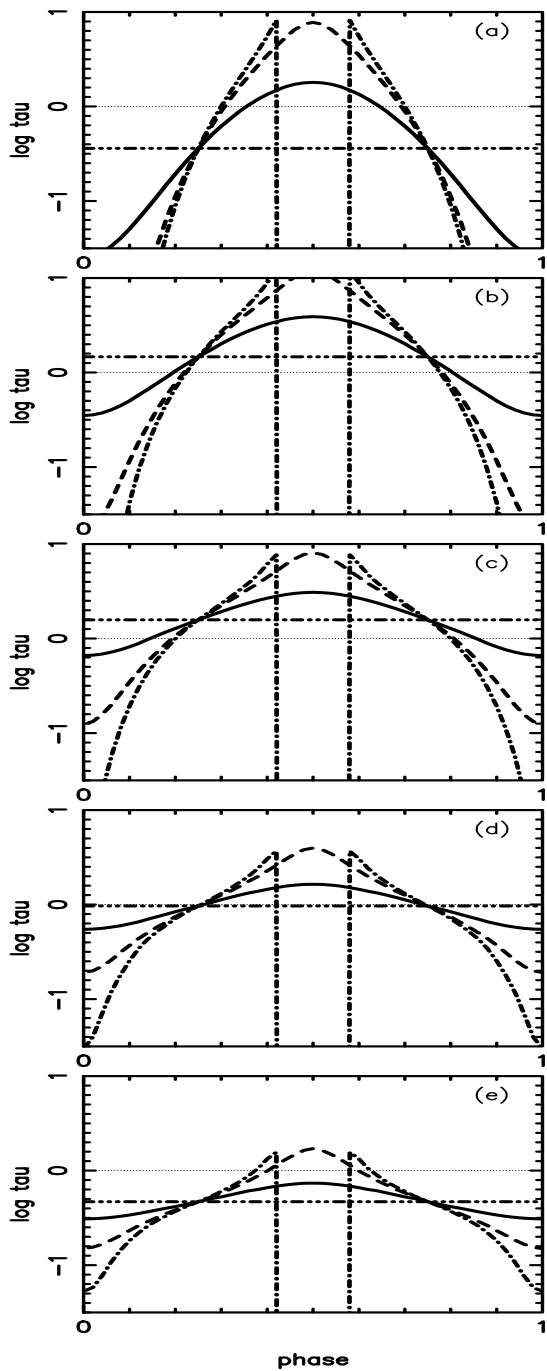

\vskip 19.truecm
\includegraphics{bpfig6a.eps}
\includegraphics{bpfig6b.eps}
\includegraphics{bpfig6c.eps}
\includegraphics{bpfig6d.eps}
\includegraphics{bpfig6e.eps}
\caption{The dependence of the $\gamma$-ray optical depths in the thermal radiation of the irradiated star as a function of the phase of the binary system. The phase is measured from the situation in which the X-ray source is in front of the star (in respect to the observer). Specific figu\-res show the $\gamma$-ray optical depth light curves for selected energies of $\gamma$-ray photons equal to: 33 GeV (a), 100 GeV (b), 330 GeV (c), 1 TeV (d),
and 3.3 TeV (e). The light curves for different inclination angles of the binary system are shown for: $\alpha = 0^{\rm o}$ (triple-dot dashed curve), $30^{\rm o}$ (solid), $45^{\rm o}$ (dashed), $60^{\rm o}$ (dashed), and $89^{\rm o}$ (dot-dashed).
The thin dotted line marks the optical depth equal to unity. 
The parameters of the star are following: $R_\star = R_\odot$ and its surface temperature $T_\star = 6000$ K. The orbit of the X-ray source is circular 
with the radius $R_{\rm b} = 2R_\star$. The $\gamma$-rays are injected from the vicinity of the X-ray source.}
\label{fig6}
\end{figure}

We investigate also how the $\gamma$-ray optical depths can change in the case of a low mass compact binary system in which the X-ray source is on the circular orbit around the solar type star.
In this case the angle $\beta$ between the direction towards the observer and the
direction defined by the X-ray source and the star can change significantly 
depending on the inclination angle $\alpha$ of the binary system. 
$\beta$ is determined in this case by the phase $\Phi$ of the X-ray source
(for geometrical situation see Fig.~5).

The example $\gamma$-ray optical depth light curves are calculated for the range of the inclination angles of the binary system and specific energies of $\gamma$-ray photons injected in the direction towards the observer. The results for the X-ray source on a circular orbit with the radius $R_{\rm b} = 2R_\star$ and the injection place of $\gamma$-rays also at the X-ray source are shown in Fig.~6. Note the characteristic optical depth light curves in the case of the non-eclipsing systems with the inclination angles different than zero. In such case,
the optical depths reach the maximum when the X-ray source (and also $\gamma$-ray source) is behind the star (in respect to the observer). The optical depths at the peak of the maximum are larger for larger inclination angles of the binary system. For the considered parameters of the binary system, the optical depths are above unity at least for some range of phases close to the phase $\Phi = 0.5$.
In the case of the eclipsing binaries the observer should see two peaks in the optical depth light curve due to the eclipse of the X-ray and $\gamma$-ray source by the star. 
Note, that at the phases at which the $\gamma$-ray optical depth light curves reach the maximum, the hot, irradiated part of the stellar surface is not well visible. Therefore, it is natural to expect unticorrelation between the maximum of $\gamma$-ray light curve and the maximum in the optical light curve from such low mass binary systems. This prediction might be useful 
in order to increase the probability of the detection of the GeV-TeV $\gamma$-ray signal from the compact X-ray luminous LMXBs.

Since the assumed orbit of the X-ray source (and $\gamma$-ray source) is circular, the light curve looks symmetric. This light curve can be significantly modified when the distance of the X-ray source changes with the phase of the binary.
In order to have impression about the influence of the ellipticity of the orbit, we calculate the optical depth light curves for other distances of the X-ray and $\gamma$-ray sources from the star.  In Fig.~7, we investigate the optical depths for the orbits of the X-ray source in the range $R_{\rm b} = 2-3R_\star$ and distances of the $\gamma$-rays source in the range $R = 1.5-3R_\star$. Note significant change of the 
optical depths with these two basic parameters $R_{\rm b}$ and $R$. However, 
for the whole range of these parameters, the optical depths are always above unity
for some range of phases. Therefore, we expect that intensive absorption (and also production) of $\gamma$-ray photons might occur in the case of compact low mass X-ray
binary systems. However in the realistic case, the spectrum of $\gamma$-ray photons should be formed in the complicated cascade process due to 
a highly anisotropic soft radiation created by such irradiated star.
This cascade scenario will be clearly more complicated than considered up to now
for the $\gamma$-ray production inside the massive binary systems (see e.g. early results by Bednarek~2000).

For comparison, we also show the $\gamma$-ray light curves for the case of stellar mass 
companion which is in the giant phase (other parameters of the binary as in Fig.~7b). Due to larger distance of the X-ray source from the surface, the optical depths are clearly lower than in the case of the binary system with the solar mass star on the Main Sequence.
However, still for small range of phases and relatively large inclination angles of the binary system, the optical depths overcome unity. Therefore, in some geometrical situations also these more geometrically extended binaries (i.e. with longer orbital periods) can provide conditions for efficient production of GeV-TeV $\gamma$-rays.

Calculated here $\gamma$-ray optical depth light curves allow us to conclude that
the largest fluxes of GeV $\gamma$-rays should be expected in the phases where
the optical depths are clearly above unity. In such cases the cascading effects degrade
a part of the primary $\gamma$-ray spectrum in the TeV energy range due to efficient cascading. On the other hand, the largest fluxes of TeV $\gamma$-rays are expected in phases where the optical depths are close to unity. In such case, primary TeV $\gamma$-rays are efficiently produced and not severely absorbed in the IC $e^\pm$ pair cascades. Based on the analysis of Figs. 6 and 7, we expect the largest fluxes of the GeV $\gamma$-rays when the X-ray source (a compact object) is behind the normal star
(except the eclipsing binaries). The largest fluxes of the TeV $\gamma$-rays are expected for the intermediate phases, i.e. just before and after the largest fluxes of GeV $\gamma$-rays. 

\section{The example cases of LMXBs}

As an example, we show the  $\gamma$-ray optical depth light curves for two
LMXB systems which show optical modulation with the binary system period.
This modulation is interpreted as due to the irradiation of the companion star by the
X-ray source (e.g. Milgrom~1976). They have been reported in the past as a possible TeV-PeV $\gamma$-ray sources.

\begin{figure}[t]
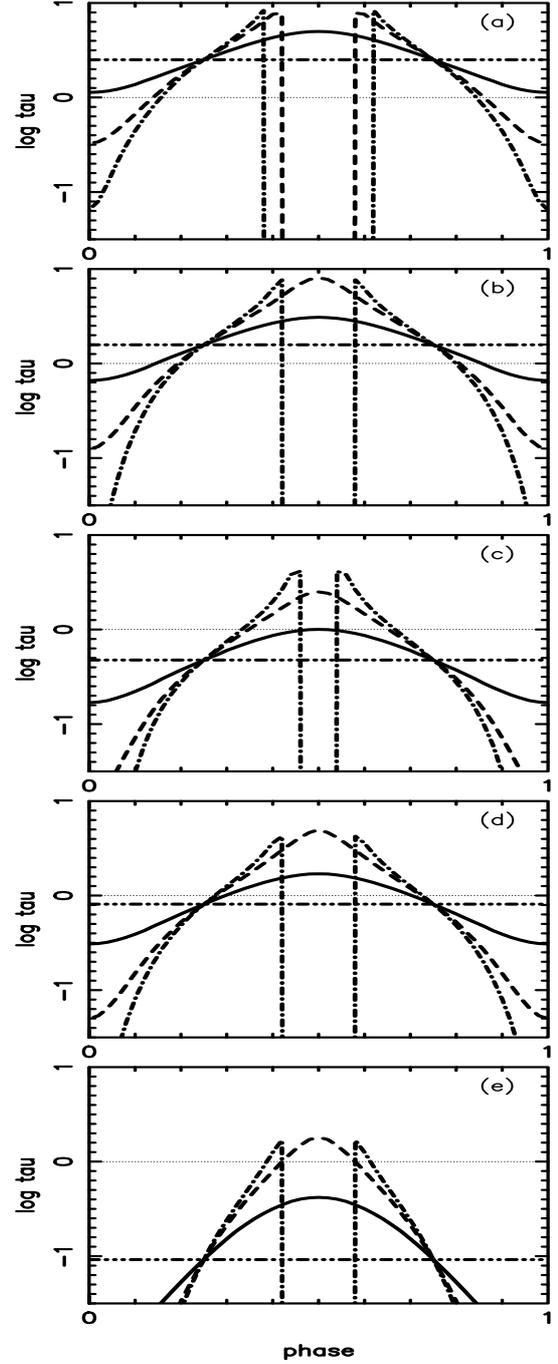

\vskip 19.truecm
\includegraphics{bpfig7a.eps}
\includegraphics{bpfig7b.eps}
\includegraphics{bpfig7c.eps}
\includegraphics{bpfig7d.eps}
\includegraphics{bpfig7e.eps}
\caption{As in Fig.~6 but for different distances of the X-ray source from the star
($R_{\rm b}$) and different injection places of $\gamma$-ray photons ($R$): $R_{\rm b} = 2R_\star$ and $R = 1.5R_\star$ (a), $R_{\rm b} = 2R_\star$ and $R = 2R_\star$ (b), $R_{\rm b} = 3R_\star$ and $R = 3R_\star$ (c), and $R_{\rm b} = 3R_\star$ and $R = 2R_\star$ (d).
The case of the solar mass star in the giant phase is considered ($R_\star = 100R_\odot$ and $T_\star = 3000$ K) for $R_{\rm b} = 2R_\star$ and $R = 2R_\star$ (see figure e). It is assumed that $\gamma$-ray injection place is always on the line defined by the X-ray source and the star. The energies of $\gamma$-ray photons have been fixed on 330 GeV.}
\label{fig7}
\end{figure}

\subsection{Her X-1}

The famous LMXB, Her X-1,
is characterised by the X-ray luminosity $L_{\rm x} = 6\times 10^{37}$ erg s$^{-1}$
(White et al.~1983). The com\-pa\-nion star has the radius $R_\star = 3.86R_\odot$, the mass $M_\star = 2M_\odot$ with a neutron star on an orbit with the radius $R_{\rm b} = 8.61R_\odot$. The inclination of the binary system has been estimated on $\alpha = 80^{+8}_{-5}$ degrees and its orbital period is 1.7 days (Nagase~1989).

Her X-1 has been claimed in the past as a pulsed TeV $\gamma$-ray source
by the Whipple Collaboration (e.g. Gorham et al.~1986, Lamb et al.~1988). However, later analysis have not confirmed these reports (Reynolds et al.~1991).
The modern Cherenkov telescope collaborations have not reported any observations of this source recently. 

We have calculated the $\gamma$-ray optical depth light curves for the parameters of Her X-1. It is assumed that the source of primary $\gamma$-rays (produced by accelerated electrons) is close to the X-ray source. The results are shown in Fig.~8 for selected energies of $\gamma$-ray photons. 
It is clear that the optical depths are greater than unity for some relatively small range of phases in the case of $\gamma$-rays with energies in the range $E_\gamma = 0.1-1$ TeV. The optical depths can reach the maximum values $\tau\sim 3$. Therefore, the possible cascading effects have to be taken into account when  calculating detailed $\gamma$-ray spectrum escaping towards the observer. However, the primary $\gamma$-ray spectrum should not be very severally degraded at the TeV $\gamma$-ray energies as expected in the case of massive binary systems such as LS 5039 or LSI 61 303.
We predict that the largest fluxes of GeV-TeV $\gamma$-rays should appear
in the range of phases just before and after the eclipse of the X-ray source by 
the  companion star, i.e. in the ranges: $\Phi\sim 0.35-0.45$ and $0.55-0.65$.
Note that zero phase corresponds to the situation where the X-ray compact source is in
front of the companion star. 
Moreover, this $\gamma$-ray emission should be unticorrelated with the optical emission coming from a part of the stellar surface irradiated by the X-ray source. In fact, evidences of the optical modulation with the period of the 
binary system Her X-1 has been clearly detected in the past.

For the injection place of primary $\gamma$-rays which
significantly differ from the location of the X-ray source (e.g. farther from the X-ray source along the jet or at the shock structure), the phase patterns of the $\gamma$-ray optical depths should of course change accordingly.

\subsection{Sco X-1}

The second binary system, Sco X-1, has significantly different parameters from Her X-1.
It contains a low mass star, $M_\star = 0.5 M_\odot$, and a neutron star (Liu et al.~2007). Sco X-1 belongs to the class of Z sources with three states of X-ray emission characterised by the luminosities  $L_{\rm x} = (4-12)\times 10^{37}$ erg s$^{-1}$ (Hasinger \& van der Klis~1989). The accretion disk around the neutron star in this binary system launches a jet. Therefore, the object belongs also to the class of microquasars. The inclination angle of the binary system has been estimated in the range: $\alpha = 44^{\rm o}\pm 6^{\rm o}$ (Fomalont et al.~2001). The period of the binary system is 18,9 hrs (Gottlieb et al.~1975). 
In our calculations we apply the radius of the companion star $R_\star = 1R_\odot$
(evolved subgiant, Steeghys \& Casares~2002),
the radius of the binary system $R_{\rm b} = 2.14r_\odot = 1.5\times 10^{11}$ cm, the inclination of the binary system $\alpha = 44^{\rm o}$, and the power of the X-ray source $10^{38}$ erg s$^{-1}$.

Sco X-1 has been also claimed as a TeV-PeV $\gamma$-rays source in the 80-ties but with the low significance  (e.g. Brazier et al.~1990, Tonwar et al.~1991). Also these results have not been confirmed by latter observations.

As for Her X-1, the $\gamma$-ray optical light curves have been calculated as a function of energy of $\gamma$-ray photons. Note, that Sco X-1 is  more compact binary system than Her X-1. Therefore, the range of phases, for which the $\gamma$-ray optical depths are above unity, is significantly broader. Also, the absolute values of the optical depths are larger. So then, we expect much stronger modification of the escaping $\gamma$-ray spectra by IC $e^\pm$ pair cascading effects.
Due to these reasons, the TeV $\gamma$-ray emission can be severally attenuated for the 
phases close to $\sim 0.5$ and the TeV $\gamma$-ray light curve may show two minima,
first due to inefficient production in the region close to $\sim 0.0$ phase and the second due to strong absorption in the region close to the phase $\sim 0.5$.
In contrast, the GeV $\gamma$-ray light should show one broad maximum centered on
the phase $0.5$. These conclusions are derived assuming the circular orbit of the
compact object around the companion star. The more detailed analysis is at present not
possible due to the lack of information on the details of the orbit of the compact object in Sco X-1 (i.e. ellipticity, phase of periastron). 

It is not clear at present whether the companion star in Sco X-1 is really efficiently irradiated by the X-ray source. Early calculations by Milgrom~(1976) suggest that irradiation is
important and consistent with observational results at that time. In these calculations relatively small inclination angles has been considered (up to $i\sim 6^o$). The analysis of more recent observations turns to rather intermediate inclination angles (Fomalont et al.~2001,Steeghys \& Casares~2002) which is applied in this paper. The present lack of strong modulation of UV emission
might be understood assuming that significant part of the hot region on the stellar surface
is shadowed by the large accretion disk in this system. Note, that the outer radius of this accretion disk has been estimated on $\sim 6\times 10^{10}$ cm (Vrtilek et al.~1991) which is comparable to the separation of stars in this binary.

\section{Discussion and Conclusion}

\begin{figure}[t]
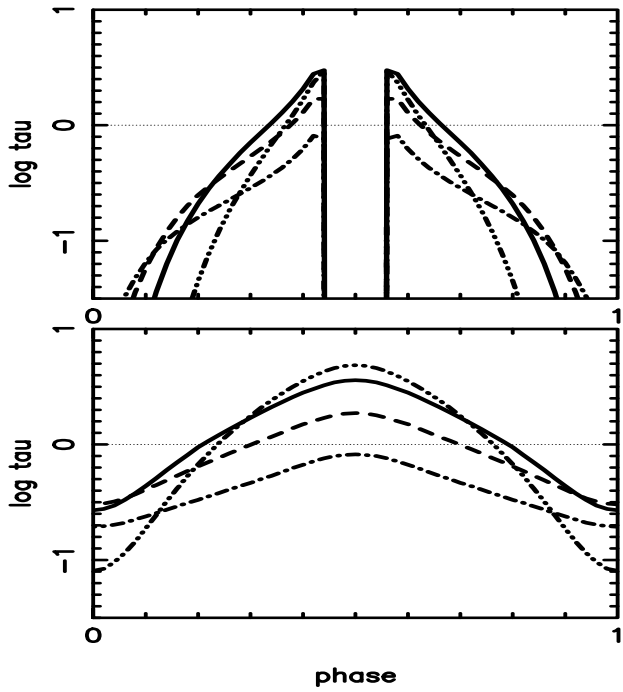

\vskip 10.truecm
\includegraphics{bpfig8a.eps}
\includegraphics{bpfig8b.eps}
\caption{The $\gamma$-ray optical depth light curves for two binary systems with the parameters mentioned in the main text: Her X-1 (upper figure) and Sco X-1 (bottom). 
Specific curves show results for different energies
of $\gamma$-ray photons: 100 GeV (triple-dot-dashed curve), 330 GeV (solid), 1 TeV (dashed), and 3.3 TeV (dot-dashed). It is assumed that the injection place of the $\gamma$-rays is at the location of the X-ray source.}
\label{fig8}
\end{figure}

We showed that the optical depths for TeV $\gamma$-rays, injected inside the LMXBs close to the surface of the companion stars irradiated by the X-ray sources, are larger than unity for specific locations of the observer. It is easy to show that  
due to similar cross section for the $\gamma-\gamma\rightarrow e^\pm$ absorption process and the cross section for the ICS process of soft photons by relativistic electrons, also the optical depths for electrons on these same stellar radiation should have similar values. Therefore, it is expected that the production and cascading
of $\gamma$-rays can be efficient in the LMXBs in which the companion star is effectively heated by the close X-ray source. However, the optical depths in the case of irradiated low mass stars are generally lower than the optical depths expected in the HMXBs detected in the TeV $\gamma$-rays (such as LS 5039 or LSI 303 +61, e.g. Bednarek~2006a). So then, the cascading effects in the LMXBs should not occur so efficiently as it was shown 
for those sources (Bednarek~2006b,~2007, Sierpowska-Bartosik \& Torres~2007,~2009). 
Due to these substantial optical depths, we expect that the GeV-TeV $\gamma$-ray emission can be also efficiently produced within compact LMXBs in which the compact X-ray source strongly irradiates companion star provided that electrons are accelerated to TeV energies also inside LMXBs. 
The mechanism of particle acceleration inside LMXBs can either differ from that one operating within
the HMXBs or can be quite similar. In fact, the TeV $\gamma$-rays are observed from  different types of sources (e.g. jets of AGNs, vicinity of pulsars, or shocks of supernova remnants). Therefore,
different conditions around compact objects in HMXBs and LMXBs can also turn to acceleration of 
particles to TeV energies. On the other hand, TeV $\gamma$-ray emission inside LMXBs may  
be produced in similar scenarios as proposed for HMXBs (e.g. accreting neutron star scenario
proposed recently by Bednarek~2009b).
However, as we have shown above the TeV $\gamma$-ray emission from LMXBs should be limited to specific narrow range of phases which are expected to be different from the range of phases observed in the HMXBs due to differences in the soft radiation field (see considered example cases of Her X-1 and Sco X-1).

Based on the results of calculations of the optical depths and a simple application of the law of gravity, we can envisage which LMXBs are expected to be potential GeV-TeV
$\gamma$-ray sources.
We conclude that in the case of the solar type star in the evolutionary phase on the Main Sequence ($R_\star = R_\odot$ and $M_\star = M_\odot$) 
and the orbit with the radius $R_{\rm b} < 3R_\star$, the $\gamma$-rays have a chance to be efficiently absorbed which means also their efficient production in IC process by electrons. This will occur 
provided that the period of the binary system is shorter than $P_{\rm b}\sim 17 R_\star^{3/2}M_\star^{-1/2}$ hrs, where $R_\star$ and $M_\star$ are in the solar units. Note that many LMXBs have the orbital periods within this limit (Liu et al.~2007).

Based on the analysis of the optical depth light curves we conclude that
the emission patterns of escaping $\gamma$-rays from the low mass binary system 
strongly depend on its inclination angle. We predict that in the simplest considered model of a point like source of primary $\gamma$-rays on a circular orbit around irradiated low mass star, the observer located at:

\begin{itemize}

\item  {\it Small inclination angles}: detects significant 
GeV-TeV $\gamma$-ray emission only from very compact binaries in which the compact object is at the distance of less than 2 stellar radii from the companion star
(compare the triple-dot dashed lines with the thin dotted lines in Figs.~6 and 7).
In this case the modulation of the UV emission from an irradiated companion star with the period of the binary system may be significantly reduced since the hottest parts of the stellar surface are not well visible.

\item {\it Intermediate inclination angles}: detects a single pulse of $\gamma$-ray emission with the maximum corresponding to the phases when the X-ray source is behind the companion star. 
In the case the observer should detect significant modulation of the UV emission from the 
stellar surface modulated with the period of the binary system.

\item {\it Large inclination angles and eclipsing binaries}: detects two narrow pulses of $\gamma$-rays just before and after the eclipse of the X-ray source by the companion star. 
In this case the modulation of the UV emission should be clearly observed but in some cases can be also strongly suppressed due to the obscuration of the hottest parts on the stellar surface by the large accretion disk.

\end{itemize}

The $\gamma$-ray emission features from LMXBs can differ from those observed in HMXBs. The differences  are due to the fact that in general the optical depths for $\gamma$-rays 
in the radiation field of irradiated stars in LMXBs are lower in respect to these calculated in the case of HMXBs. 
We want to mention that the specific binary system may in fact differ significantly
from the simplified picture discussed in this paper. The main geometrical complications can be introduced by the inhomogeneous irradiation of the star by X-ray source (anisotropic X-ray source) and the elliptic orbit of the X-ray source around the companion star. Moreover, our analysis base on the assumption that the possible
cascade initiated by primary electrons and $\gamma$-rays develop mono-directionally
through the binary system. In fact, the structure of the magnetic field of the star can  
influence the paths of secondary cascade $e^\pm$ pairs. As a result, the escaping secondary $\gamma$-rays can form complicated structures on the sky (see Sierpowska
\& Bednarek~2005).

Someone can wonder whether absorption of $\gamma$-rays should not occur efficiently in the radiation field of the X-ray source on the surface of the accreting neutron star. We show below that this is not the case provided that $\gamma$-rays are injected at some reasonable distance from the X-ray source.
Assuming the characteristic dimension of the X-ray source equal to $\sim 10^5$ cm (the order of the polar cap on the neutron star surface) and the luminosity of the X-ray source $10^{38}$ erg s$^{-1}$, we estimate the characteristic temperature of the X-ray source on $T_{\rm X}\approx  6\times 10^7 L_{38}^{1/4}R_5^{-1/2}$ K, where $R_{\rm x} = 10^5R_5$ cm is the characteristic dimension of the X-ray source.  Note, that for simplicity we assumed the black body type emission from the X-ray source.
We estimate the distance, $D$, from such X-ray source at which the optical depths in the radiation of such X-ray source become lower than unity from
$\tau = D\sigma_{\gamma\gamma\rightarrow e^\pm}n_{\rm x}$, where
$\sigma_{\gamma-\gamma\rightarrow e^\pm}$ is the cross section for $e^\pm$ pair creation in $\gamma-\gamma$ collision, $n_{\rm x}\approx 20T^3(R_x/D)^2$ cm$^{-3}$ is density of X-ray photons at the distance $D$. Applying that 
the cross section for $\gamma-\gamma$ collision is inversely proportional to energy of $\gamma$-ray photon, we estimate that the optical depth becomes lower than unity
for distances larger than $D\sim 10^8$ cm and energies of $\gamma$-rays above $\sim 10$ GeV.  
Note moreover that most of the $\gamma$-rays move in the outward direction in respect to the location of the X-ray source. Thus, the geometrical effects can additionally decrease the optical depths in the X-rays. This estimate on the possible location of the $\gamma$-ray source is clearly consistent with the dimensions of the binary system. 

Based on similar estimations it is possible to show that the radiation field from the inner part of the accretion disk around the neutron star may not prevent escape of $\gamma$-ray photons
even if they are created within or close to this region. For example, in the case of Her X-1 the inner disk 
temperature is estimated in the range $\sim (1.8-2.5\times 10^4)$ K (Sazonov~2009) and the disk inner radius is at  the distance of the order of a few $\sim 10^8$ cm from the neutron star (Ghosh \& Lamb~1979). We estimate the mean free path for $\gamma$-rays propagating in such radiation on
$\sim 5\times 10^9$ cm. This is clearly larger than dimension of the disk inner radius.
Therefore, $\gamma$-rays can escape from the inner part of the accretion disk without significant absorption. 
Note, that the region of particle acceleration may be located at some distance from the hot neutron star surface and the inner accretion disk but still inside the binary systems. For example, there are models which propose acceleration of particles within the jets which are observed in the case of some LMXBs (e.g. Sco X-1).

We can envisage realistic situation in which the absorption of TeV $\gamma$-rays in the radiation of the X-ray source can be safely neglected. 
For example, in the case of Her X-1, the X-ray source can be produced on the neutron star surface due to accretion of matter. However, the acceleration of electrons (and production of the primary $\gamma$-rays) can occur close to or in a very turbulent, magnetized region at the inner accretion disk radius where the pressure of the accreting matter is balanced by strongly magnetized rotating neutron star magnetosphere (the surface magnetic field of Her X-1 is estimated on $2.9\times 10^{12}$ G, e.g. Karino~2007). Then, the inner radius of the accretion disk in Her X-1 is expected at the distance of a few $10^8$ cm from the neutron star. In the case of Sco X-1
production of $\gamma$-rays can occur inside a part of the jet at some distance from the 
neutron star and the inner disk but still inside the binary system.

\begin{acknowledgements}
This work is supported by the Polish MNiSzW grant N N203 390834. 
\end{acknowledgements}


\end{document}